\documentclass[prl,twocolumn,floatfix,showpacs,showkeys,preprintnumbers,amssymb,amsmath]{revtex4}
\usepackage{graphicx}
\usepackage{dcolumn}
\usepackage{bm}
\usepackage[mathscr]{eucal}

\begin{document}
\newcommand{\beq}{\begin{equation}}
\newcommand{\eeq}{\end{equation}}
\newcommand{\ket}{\rangle}
\newcommand{\bra}{\langle}
\newcommand{\A}{\mathbf{A}}
\preprint{ }
\title{Efficient decomposition of quantum gates}
\author{Juha J. Vartiainen}
\email{juhav@focus.hut.fi}
\author{Mikko M\"ott\"onen}
\author{Martti\ M.\ Salomaa}
\affiliation{Materials Physics Laboratory, POB 2200 (Technical Physics), FIN-02015 HUT,
Helsinki University of Technology, Finland}

\date{\today}

\begin{abstract}
Optimal implementation of quantum gates is crucial for designing a
quantum computer. We consider the matrix representation of an
arbitrary multiqubit gate. By ordering the basis vectors using the
Gray code, we construct the quantum circuit which is optimal in
the sense of fully controlled single-qubit gates and yet is equivalent
with the multiqubit gate. In the second step of the optimization,
superfluous control bits are eliminated, which eventually results
in a smaller total number of the elementary gates. In our scheme
the number of controlled NOT gates is $O(4^n)$ which coincides with the
theoretical lower bound.
\end{abstract}


\maketitle

Since the early proposal of a quantum-mechanical
computer~\cite{Feynman}, quantum superposition and entanglement
has been discovered to be potentially useful for computing.
For example, Shor's integer factorization \cite{Shor97} and
Grover's database search \cite{Grover} show considerable speed-up
compared to the known classical algorithms. Moreover, the
framework of quantum computing can be used to describe
intriguing entanglement-related phenomena, such as quantum teleportation and
quantum cryptography.

Quantum circuits~\cite{Deutsch} provide a method to implement an
arbitrary quantum algorithm. The building blocks of quantum
circuits are quantum gates, i.e., unitary transformations acting on a set of qubits.
 It has previously been shown that a
general quantum gate can be simulated exactly
\cite{elementary,divincenzo,lloyd} or approximately
\cite{kitaev,Knill} using a quantum circuit built of elementary
gates which operate only on one and two qubits. Some individual
gates operating on $n$ qubits, such as the quantum Fourier
transform, reduce to a polynomial number of elementary gates in
$n$. Unfortunately, this is not the case for an arbitrary
$n$-qubit gate, i.e., a unitary operation having $4^{n}$ degrees
of freedom. From the practical point of view, the maximum coherent
operation time of the quantum computer is limited by undesirable
interactions with the environment, i.e., decoherence. On the other hand, the number of the elementary gates involved in the decomposition governs the execution time of the quantum algorithm.
 Hence the
complexity of these quantum-circuit constructions
 is of great
interest.

The conventional approach of reducing an arbitrary $n$-qubit gate
into elementary gates is given in Ref.~\cite{elementary} and
studied with the help of examples in Refs.~\cite{Nielsen,cybenko}.
The main idea is to decompose the unitary matrix $U$, which
represents the quantum gate, into two-level matrices and to find a
sequence of C$^{n-1}V$ and C$^{n-1}$NOT gates which implements
each of them. Here we refer with C$^{k}V$ to the one-qubit gate
$V$ having $k$ control bits. The control bits, each of which has
the value zero or one, specify the subspace in which the gate $V$
operates. This $2^{n-k}$-dimensional subspace consists of those
basis vectors for which the values of the controlled qubits match
with those of the control bits. In this approach, a number of
C$^{n-1}$NOT gates is required to change the computational basis,
such that the two-level matrix under consideration represents the
desired C$^{n-1}V$ gate.

For the purpose of their physical implementation, all the
C$^{n-1}V$ gates can be further decomposed into a sequence of
elementary gates, for instance, using the quantum circuit of
Ref.~\cite{elementary}. For the simulation of a C$^{n-1}$NOT or
C$^{n-1}V$ gate, a quantum circuit of $O(n^2)$ elementary gates is
required while C$^{n-1}W$ requires only $O(n)$ gates, provided
that $W$ is unimodular. In Ref.~\cite{elementary}, it was
considered that since $O(n)$ C$^{n-1}$NOT gates are
needed between each of the $O(4^n)$ C$^{n-1}V$ gates, the total
circuit complexity is $O(n^34^n)$. It has recently been shown with
the help of palindromic optimization \cite{Aho}, that the number
of C$^{n-1}$NOT gates required in the simulation can be reduced to
$O(4^n)$ which results in circuit complexity $O(n^{2}4^n)$. A constructive upper bound for the optimal
circuit complexity has been reported~\cite{Knill} to be $O(n4^n)$ \cite{diag}
which may also be achieved by combining the previous results
\cite{elementary,Aho} with the fact that C$^{n-1}$NOT gates may be
replaced with proper CNOT gates upon changing the computational
basis \cite{cybenko}. The theoretical lower bound~\cite{Shende}
for the number of CNOT gates needed to simulate an arbitrary
quantum gate is $\lceil (4^n-3n-1)/4\rceil$. However, no circuit
construction yielding a complexity less than $O(n4^n)$ has been
reported, nor could be trivially combined from the previous
results.

In this Letter, we show how to construct a quantum circuit
equivalent to an arbitrary $n$-qubit gate. The circuit obtained
has complexity $O(4^n)$ which scales according to the predicted
theoretical lower bound. The scheme utilizes the reordering of
the basis vectors, i.e., instead of labeling the basis vectors
through the binary coding, we rather employ Gray
codes~\cite{numres}. The special property of any Gray code basis
(GCB) is that only one bit changes between the adjacent basis vectors.
Hence no C$^{n-1}$NOT gates are needed in the decomposition.
Furthermore, we find that only a small fraction of the control
bits appears to be essential for the final result of the
decomposition. Finally, the further elimination of futile control
bits reduces the circuit complexity from $O(n4^n)$ down to
$O(4^n)$.

The physical state of an $n$-qubit quantum register can be
represented with a vector $|\Phi\ket$ in the associated Hilbert
space $\mathbb{C}^{N}$, where $N=2^n$. In a given basis
$\{|e_k\ket\}$, a quantum gate acting on a $n$-qubit register corresponds to a
certain $2^n \times 2^n$ unitary matrix $U$.
The QR-factorization of
any matrix can be performed using the Givens rotation matrices~\cite{Golub}. A
Givens rotation $^iG_{j,k}$ is a two-level matrix which operates
non-trivially only on two basis vectors, $|e_j\ket$ and
$|e_k\ket$. We
define $^iG_{j,k}=G_{j,k}(A)$ to be a generic rotation matrix
which selectively nullifies the element on the $i^{\rm th}$ column
and the $j^{\rm th}$ row with the help of the element on the $i^{\rm th}$
column and the $k^{\rm th}$ row of a matrix $A$. The nontrivial
elements of the two-level matrix
$^iG_{j,k}=\{^ig_{l,n}\}_{l,n=1}^N$ acting on the matrix
$A=\{a_{l,n}\}_{l,n=1}^N$ are given by
$$
^i\Gamma_{j,k}:= \!\!
\left(%
\begin{array}{cc}
^ig_{k,k} & \!^ig_{k,j}  \\
^ig_{j,k} & \!^ig_{j,j}  \\
\end{array}\!\!
\right)\!
=\frac{1}{\sqrt{|a_{j,i}|^2\!+|a_{k,i}|^2}}\left(%
\begin{array}{cc}
a^*_{k,i} & \!a^*_{j,i}  \\
\!\!-a_{j,i} & \!a_{k,i}  \\
\end{array}%
\right),
$$
while the other elements match with the identity matrix. In the special case where the element $a_{j,i}$ vanishes, the Gives rotation is defined to be an identity matrix.

For example, the first Givens rotation we employ results in
$$
{}^1G_{N,N-1} U= \!\left(%
\begin{array}{cccc}
u_{1,1} & u_{1,2} & \ldots & u_{1,N}  \\
\vdots & \vdots & \ddots &\vdots  \\
u_{N-2,1} & u_{N-2,2} & \ldots & u_{N-2,N}  \\
\tilde{u}_{N-1,1} & \tilde{u}_{N-1,2} & \ldots & \tilde{u}_{N-1,N}  \\
0 & \tilde{u}_{N,2} & \ldots & \tilde{u}_{N,N}  \\
\end{array}%
\right),
$$
where the modified elements of $U$ due to
${}^1G_{N,N-1}$ are indicated with the tilde. Applying
${}^1G_{N-1,N-2}$ to the modified matrix we can nullify the element
$\tilde{u}_{N-1,1}$ and similarly the whole first column, except
the diagonal element. The definition of the Givens rotation
ensures that the argument of the diagonal element vanishes and the
unitarity of the matrix $U$ fixes its absolute value to unity. The
process is continued through the columns $2$ to $N-1$, resulting
in an identity matrix, except for the diagonal element on the
$N^{\rm th}$ row which becomes $\det(U)$. In fact, without loss of
generality we may assume that  $U\in SU(2^n)$,
since the nonzero argument of the determinant of $U$ contributes
only to the global phase of the state vector $|\Phi\rangle$ which is
not measurable. Thus we obtain the factorization
\begin{eqnarray}\label{eq:gu}
\left(\prod_{i=1}^{2^n-1}\prod_{j=i+1}^{2^n} {^{2^n-i}}G_{j,j-1}\right)U= I,
\end{eqnarray}
where the order of the products is taken from left to right, i.e., the first element $^{2^n-1}G_{2^n,2^n-1}$ is the leftmost matrix in the product. The assumption of the unimodularity of the matrix $U$ may be
dropped if one first applies a matrix $e^{-i\arg[\det(U)]/N}I$ which may be realized with a single one-qubit gate.

\begin{figure}
\includegraphics[width=0.45\textwidth]{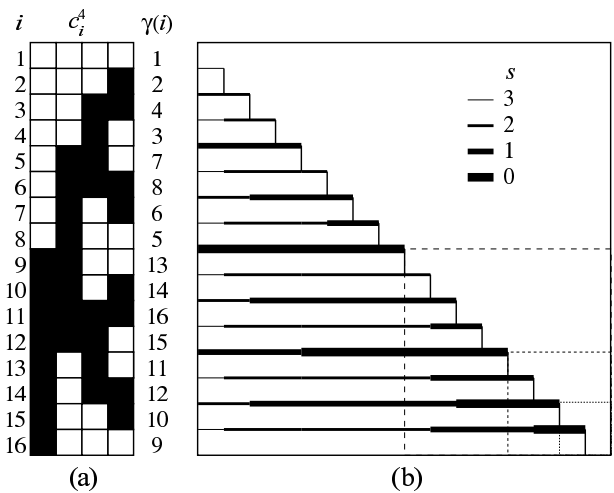}
\caption{\label{fg:peli}
(a) Illustration of the Gray code $c^4_i$. White squares stand for bit values $0$ and black squares denote $1$. The function $\gamma(i)$ represents the value of the bit string $c^4_i$ plus one.
(b) Table of dimensions $2^4\times 2^4$ shows the number of control bits used while nullifying the elements of the matrix $U$. The width of the line $L^s$ represents the number $p=3-s$ of control bits required to zero the element below the line with the element above it.}
\end{figure}

For quantum computation, it is convenient to choose the basis
vectors according to $|e_k\rangle=\bigotimes_i |x_i^k\rangle$, where
$x_i^k \in \{0,1\}$ and the index $i=1,...,n$ refers to the
physical qubit $i$.  Here we note that the order of the basis
vectors in the computational basis is not fixed. In the previous
approaches \cite{elementary,Aho}, the order of the basis
vectors has been chosen such that the values $x_i^k$ essentially
form the binary representation of the number $k-1$, i.e.
$k=1+\sum_{i=0}^n 2^i x_i^k$. However, the coefficients $x_i^k$
can just as well be chosen to form a Gray code \cite{numres}
corresponding to the number $k-1$. A Gray code of $n$ qubits
$\{c^n_1,c^n_2,\dots,c^n_{2^n}\}$ is a palindrome-like ordering of
binary numbers having the special property that the adjacent
elements $c^n_i$ and $c^n_{i+1}$ differ only in one bit from each
other. We choose to use such a Gray code in which each bit string $c^n_i=b_n^i \cdots b_{2}^i b_1^i$ is obtained from
the binary representation $i_{\rm b}$ of the number $i$ as
$c^n_i=i_{\rm b} \, {\rm XOR} \, (i_{\rm b}/2)$.
 Furthermore, we define a function
$\gamma(i)$ to represent the value of the bit string $c^n_i$ plus
one, i.e., $\gamma(i)=1+\sum_{l=1}^{n}b_l^n 2^l$.
An example of the Gray code
and the function $\gamma$ for the case $n=4$ is presented in
Fig.~\ref{fg:peli}(a).

The advantage of using the GCB instead of the binary code basis
(BCB) is that a unitary two-level matrix operating on adjacent
basis vectors equals the matrix representation of some C$^{n-1}V$
gate. Consequently, each of the $2^{n-1}(2^n-1)$ Givens rotations
${^i}G_{j,j-1}$ can be implemented using only one fully controlled
single-qubit gate C$^{n-1}V$ and no C$^{n-1}$NOT gates are needed,
unlike in previous schemes \cite{elementary,Aho}.

\begin{figure*}
\includegraphics[width=0.92\textwidth]{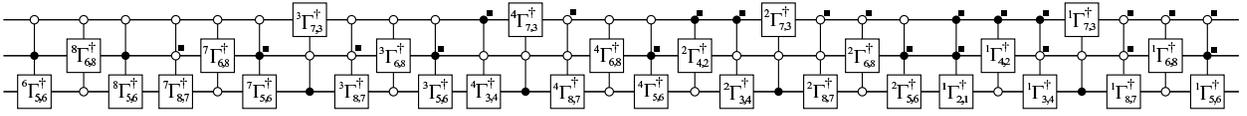}
\caption{\label{fg:3qubitfull} Quantum circuit equivalent to an arbitrary three-qubit quantum gate up to a global phase. The control bits indicated with a black square on the upper right hand side corner are superfluous and may be omitted to decrease the complexity of the decomposition, while the generic nature of the C$^k(^i\Gamma_{j,k})$ gates assures that the result remains invariant.}
\end{figure*}

Let us denote the permutation matrix accomplishing the transformation of basis from the GCB to the BCB by $\Pi$.
Since the conventional basis for the matrix representations is the BCB we rewrite
Eq.~(\ref{eq:gu}) in the BCB as
\begin{equation}\label{eq:BCBgu}
\left(\prod_{i=1}^{2^n-1}\prod_{j=i+1}^{2^n}\Pi\,\left({^{2^n-i}}G_{j,j-1}^{\rm GCB}\right)\,\Pi^\dagger\right)U^{\rm BCB}=I.
\end{equation}
Since the matrix $\Pi$ is just a permutation of the basis vectors defined by the function $\gamma$, Eq.~(\ref{eq:BCBgu}) yields
\begin{equation}\label{eq:BCBgug}
\left(\prod_{i=1}^{2^n-1}\prod_{j=i+1}^{2^n}{^{\gamma(2^n-i)}}G_{\gamma(j),\gamma(j-1)}^{\rm BCB}\right)U^{\rm BCB}=I.
\end{equation}
It is seen from
Eq.~(\ref{eq:BCBgug}) that every Givens rotation ${^{\gamma(i)}}G_{\gamma(j),\gamma(j-1)}^{\rm BCB}$ acts nontrivially only on the basis
vectors, $|e_{\gamma(j)}\rangle$ and  $|e_{\gamma(j-1)}\rangle$, for which the binary representations differ only in one bit. It is also noted that the column order of the diagonalization process is changed according to the function $\gamma$, which was not utilized in Ref.~\cite{Aho}, in which a fixed column order was assumed in the palindromic optimization. The decomposition of an
arbitrary matrix $U$ in terms of fully controlled single-qubit gates may now be constructed straightforwardly
according to Eq.~(\ref{eq:BCBgug}). It also determines the numerical values of the generic Givens rotation matrices. The quantum circuit for an arbitrary three-qubit gate is shown in Fig.~\ref{fg:3qubitfull}, where we have
assumed that all the matrices are given in the BCB.
 Since $^i\Gamma_{j,k}\in SU(2)$, the gates C$^{n-1}(^i\Gamma_{j,k})$ decompose into $O(n)$ elementary gates. Thus the gate complexity of this construction is $O(n4^n)$ which already realizes the former upper bound by Knill~\cite{Knill}.

Let the unitary matrix $U$ be given in the GCB as well as the generic Givens rotation matrices $^iG_{j,k}$,
which may be realized with a C$^{n-1}({^i}\Gamma_{j,k})$ gate. Since only matrices with consecutive indices are needed in the diagonalization procedure, we simplify the notation into $^iG_{j}:=^iG_{j,j-1}$ and $^i\Gamma_{j}:=^i\Gamma_{j,j-1}$. If $s$ control bits are removed from
a C$^{n-1}({^i}\Gamma_{j})$ gate, the matrix representation $^iG_{j}^s$ of such an operation is no more
two-level, but rather $2^{s+1}$-level, i.e., the matrix $^iG_{j}^s$ operates with the matrix $^i\Gamma_{j}$
to all pairs of basis vectors which satisfy the remaining control conditions and differ in the same bit $b^i_{m_j}$ as the bit strings $c^n_j$ and $c^n_{j-1}$. Note that the structure of the Gray code assures that the bit strings $c^n_j$ and $c^n_{j-1}$ differ in no other bits, except the bit number $m_j$.
Our aim is to diagonalize the matrix $U$ by $p$ times controlled single-qubit gates C$^p({^i}\Gamma_{j})$ in the
above given order using the minimum number of control bits. Once some element becomes zero in the
diagonalization process, we must use control bits in such a way that it does not mix with the non-zero elements.

Let us consider, for example, the diagonalization of an arbitrary four-qubit gate, for which the Gray code is shown in Fig.~\ref{fg:peli}(a). When we are about to perform the first rotation $^1G_{16}$, we may discard all the control bits from C$^{3}({^1}\Gamma_{16})$
and the matrix representation of $^1G_{16}^3$ becomes $2\times 2$-block diagonal. In the implementation of $^1G_{15}^s$, we must control
 the bit number $1$, since otherwise the matrix ${^1}\Gamma_{15}$ would operate on elements in the rows $13$ and $16$
 which is forbidden since the nonzero element on row $13$ would mix with the annihilated element on row $16$.
In the next step, where we zero
 the element on row and column $(14,1)$, we may again discard all of the control bits, since both elements in the pair $\{(15,1),(16,1)\}$
 are zero and unaffected by the action of the matrix ${^1}\Gamma_{14}$, while all the other pairs are nonzero and thus allowed to mix with each
 other. Actually, while adjusting the element in position $(j,1)$ to zero, we do not have to use upper controls, i.e., no control bits with number greater than $m_{j}$ are needed.
 When working on the second column, we may remove all the upper controls with the restriction that at least one of the
 control bits must have the value 1, since the only non-zero element in the first column at position $(1,1)$ is not allowed to mix with any other element.
To support the determination
of the control bits required, we produced Fig.~\ref{fg:peli}(b) which shows the number $p$ of control bits needed for each C$^p({^i}\Gamma_{j})$ gate in the whole diagonalization process of the matrix $U\, \in \, SU(2^4)$.

Let us assume that we are diagonalizing an arbitrary matrix $U\in SU(N)$ and aim to annihilate the element in position $(j,i)$. Provided that $j> 2^{n-1}$ and $i\leq 2^{n-1}$, all the upper controls may be dropped except that if $i-1\geq 2^{m_{j}-1}$, the bit $n$ with value $1$ is also controlled.
The number of the control bits becomes $C_{m_{j}}^i=m_{j}-1+\Theta[i-1-2^{m_{j}-1}]$, where the function $\Theta(x)=1$ for $x\geq0$ and $\Theta(x)=0$ for $x<0$. Let us denote by $g_n^0(k)$ the number of C$^kV$ gates needed while nullifying the bottom left-hand-side quarter of the matrix $U$ and similarly $g_n(k)$ for the whole diagonalization process. Since the bit $m$ differs in the two consecutive bit strings $c_j^n$ and $c_{j-1}^n$ in total $q_{m}=\max(2^{n-m-1},1)$ times on rows $2^{n}\leq j<2^{n-1}$, we obtain
\begin{eqnarray}\label{go}
g_n^0(k)&=&\sum_{m=1}^n\sum_{i=1}^{2^{n-1}}q_{m}^i\delta_{C_{m}^i,k} \\
 &=&\max(2^{n-2},2^k)+\Theta(k-1)(2^{2n-k-2}-2^{n-2}), \nonumber
\end{eqnarray}
where $\delta$ is the Kronecker delta.

The number of C$^kV$ gates needed in the diagonalization process of the top left-hand-side quarter of the matrix $U$ is $g_{n-1}(k)$, while $g_{n-1}(k-1)$ gates are needed for the bottom right-hand-side quarter. This yields a recursion relation $g_n(k)=g_n^0(k)+g_{n-1}(k)+g_{n-1}(k-1)$ with the conditions $g_m(0)=2^{m-1}$ and $g_m(m)=0$ for all $m\in\{1,2,\dots,n\}$. We rewrite the recursion relation as
\beq\label{recu}
g_n(n-i)=2^{i-1}+\sum_{m=i+1}^n g_m^0(m-i)+g_{m-1}(m-i).
\eeq
For $i=1$ the terms $g_{m-1}(m-1)$ vanish and the summation may be carried out with the help of Eq.~(\ref{go}) yielding $g_n(n-1)=3\cdot2^{n-1}-2$. 
The general solution of Eq.~(\ref{recu}) contains summations and combinatorial factors. Thus, it is more convenient to give a simple upper bound
\beq\label{app}
g_n(n-i)\leq 2^{n+i}.
\eeq
Equation~(\ref{app}) is satisfied when $i=1$ and it follows by induction using Eq.~(\ref{recu}) that the upper bound holds for all $i\in \{1,2,\dots,n-1\}$.

To calculate the number of elementary gates, we use the decompositions described in Ref.~\cite{elementary}.
Table~\ref{table1} shows the number of elementary gates calculated with the exact solution of Eq.~(\ref{recu}). For large $n$, the leading contribution to the number of CNOT gates is approximately $8.7\cdot 4^n$, while the upper bound from Eq.~(\ref{app}) yields approximately $11\cdot 4^n$.
\begin{table}
  \caption{\label{table1} Number of CNOT gates and the total number of
single-qubit and CNOT gates needed for the implementation of an arbitrary
$n$-qubit gate in the scheme described.}
\begin{tabular}{|c|c|c|c|c|c|c|c|c|c|c|}
  \hline
   n & 1 & 2 & 3 & 4 & 5 & 6 & 7 & 8 & 9 \\   \hline
  CNOT & 0 & 4 & 64 & 536 & 4156 & 22618 & 108760 & 486052 & 2078668 \\
\hline
  total & 1 & 14 & 136 & 980 & 7384 & 42390 & 208820 & 944280 &  4062520 \\
  \hline
\end{tabular}

\end{table}

In conclusion, we have presented a construction which provides an efficient way to implement arbitrary quantum gates. The initial circuit is optimal in the sense that no C$^{n-1}$NOT gates are needed to permute the basis vectors. Due to the structure of the gate sequence, we are entitled to eliminate a considerably large fraction of the control bits, which results in a circuit of complexity $O(4^n)$. We note that neither one of the two techniques alone, the GCB presentation nor the elimination of the control bits do not suffice to decrease the circuit complexity from $O(n4^n)$ to $O(4^n)$.

For certain physical realizations, the implementation of the
C$^{n-1}V$ gate is, in principle, straightforward and no decomposition into elementary
gates is needed~\cite{Wang}. To further optimize the design, one
could consider the possibility of utilizing some tailored multiqubit gates~\cite{lisa1,lisa2}, instead of
a set elementary gates or to use another decomposition for the matrix $U$ than the one into Givens rotations.

The authors acknowledge Dr. S. M. M. Virtanen for his constructive comments on this research.
This work is supported by the Foundation of Technology (Helsinki, Finland) and the Academy of Finland through Research Grants in Theoretical Materials Physics and in Quantum Computation.

\end{document}